\documentclass[
 reprint,
 superscriptaddress,
 bibnotes,
 amsmath,
 amssymb,
 aps,
 prb,
 citeautoscript,
 floatfix,
]{revtex4-2}
\usepackage{braket}

\usepackage[normalem]{ulem}
\usepackage{graphicx}
\usepackage{dcolumn}
\usepackage{bm}
\usepackage[%
    colorlinks=true,
    pdfborder={0 0 0},
    linkcolor= black,
    citecolor= black
]{hyperref}
\usepackage{etoolbox}
\usepackage[flushleft]{threeparttable}
\usepackage{siunitx}

\newcommand{\phiext}{$\Phi_\mathrm{ext}$}

\begin{document}
\preprint{APS/123-QED}
\title{Operation of a high-frequency, phase-slip qubit}

    \author{Cheeranjeev Purmessur}
    \affiliation{%
    Department of Physics, University of Illinois Urbana-Champaign, Urbana, IL 61801}
    \author{Kaicheung Chow}
    \affiliation{Holonyak Micro $\&$ Nanotechnology Lab, University of Illinois Urbana-Champaign, Urbana, IL 61801}
    \author{Bernard van Heck}
    \affiliation{Dipartimento di Fisica, Sapienza Università di Roma, Piazzale Aldo Moro 2, I-00185 Rome, Italy}
    \author{Angela Kou}\thanks{Corresponding author: akou@illinois.edu} 
    \affiliation{%
    Department of Physics, University of Illinois Urbana-Champaign, Urbana, IL 61801}
    \affiliation{Holonyak Micro $\&$ Nanotechnology Lab, University of Illinois Urbana-Champaign, Urbana, IL 61801}
    \affiliation{Materials Research Laboratory,
    University of Illinois at Urbana-Champaign, Urbana, IL 61801}

\date{\today}

\begin{abstract}   
Aluminum-based Josephson junctions are currently the main sources of nonlinearity for control and manipulation of superconducting qubits.  
A constriction-based junction provides an alternative source of nonlinearity that promises new types of protected qubits and the possibility of high-temperature and high-frequency operation through the use of superconductors with larger energy gaps.
Junctions made from such superconductors have been challenging to incorporate into superconducting qubits because of difficulty controlling junction parameters and have had extremely low lifetimes, which limited their utility.
Here we demonstrate that junctions made using titanium nitride (TiN) are a promising and controllable qubit platform. 
We use TiN junctions to build superconducting qubits based on quantum phase slips through the junction.
We operate the qubit at zero flux where the qubit frequency (\qty{\sim17}{GHz}) is mainly determined by the inductance of the qubit.
We perform readout and coherent control of the superconducting qubit, and measure qubit lifetimes \qty{>60}{\mu \second}. 
Finally, we demonstrate operation of the qubit at temperatures exceeding \qty{300}{mK}. 
Our results add the TiN-based junction as a tool for superconducting quantum information processing and opens avenues for new classes of superconducting qubits.
\end{abstract}

\maketitle

\section*{Introduction}

Aluminum-based Josephson junctions (JJ) are the workhorse of superconducting circuits and are the basis of a multitude of useful quantum devices such as quantum-limited amplifiers, single-photon detectors, and superconducting qubits \cite{yurke_observation_1989, macklin_nearquantum-limited_2015, walsh_josephson_2021,kjaergaard_superconducting_2020, krantz_quantum_2019}. 
In these devices, superconducting junctions are modeled as tunneling elements for Cooper pairs or as tunneling elements for flux quanta depending on their impedance environment.
The beneficial nonlinearity arising from flux quanta tunneling across the junction, a process known as a coherent quantum phase slip (QPS), can be obscured by unwanted capacitances when implemented using aluminum-based junctions.
Superconducting constrictions that allow QPS's are a promising implementation that offer the possibility of qubits with stronger anharmonicities than their Al JJ-based counterparts \cite{mooij_phase-slip_2005}, phase-slip oscillators that are highly sensitive and tunable via parametric drives \cite{hriscu_model_2011}, and a current standard based on quantum phase slips \cite{mooij_superconducting_2006}.
Despite these proposed benefits, constriction-based QPS junctions have only been used in a handful of experiments \cite{astafiev_coherent_2012, potter_controllable_2023,peltonen_coherent_2016,honigl-decrinis_capacitive_2023,peltonen_coherent_2013,belkin_formation_2015, de_graaf_charge_2018, shaikhaidarov_quantized_2022,rieger_granular_2023}. 
This is in part due to the difficulty with fabricating narrow constrictions. 
Recent experiments have started to incorporate superconducting constrictions into quantum interference devices \cite{de_graaf_charge_2018} and into circuits that could be used for metrology \cite{shaikhaidarov_quantized_2022}.

Incorporating constriction-based junctions into qubits would be particularly beneficial for quantum information processing.
First, superconducting constrictions can be used to create qubits based on superpositions of persistent-current states \cite{mooij_phase-slip_2005} but without the need for incorporating shunting capacitors into the qubit circuit.
Such qubit states are the basis for those used in the fluxonium \cite{manucharyan_fluxonium_2009}, the bifluxon qubit \cite{kalashnikov_bifluxon_2020}, and the 0-$\pi$ qubit \cite{brooks_protected_2013, gyenis_experimental_2021}. 
The smaller parasitic capacitances enabled by superconducting constrictions is useful since capacitors lead to spurious modes that limit the operating frequency range and control mechanisms of the above-mentioned qubits. 
Second, it has been proposed that building a circuit that contains a junction acting as a tunneling element for flux quanta and a junction acting as a tunneling element for Cooper pairs could lead to a robust noise-protected qubit \cite{le_doubly_2019}.
The lower shunting capacitances needed for implementing constriction-based QPS junctions significantly reduces the requirements for realizing such a circuit.
Finally, constriction-based QPS qubits can be made from single films of higher-energy gap superconductors than aluminum. 
These qubits could have transition frequencies higher than those of current superconducting qubits, which would enable qubit operation above dilution refrigerator temperatures and would ease signal transduction between superconducting qubits and higher-energy quantum communication platforms such as optical photons.
While GrAl-based qubits \cite{rieger_granular_2023} are a promising pathway to realize devices with lower capacitances, they remain limited for use at high temperatures by the transition temperature of aluminum.
It has been singularly challenging, however, to build non-Al constriction-based qubits \cite{astafiev_coherent_2012, peltonen_coherent_2016}.
Qubit frequencies could not be determined through design \cite{astafiev_coherent_2012}, and qubit coherence times were only inferred by spectroscopic linewidths \cite{peltonen_coherent_2016}. 

\begin{figure}
\centering
\includegraphics{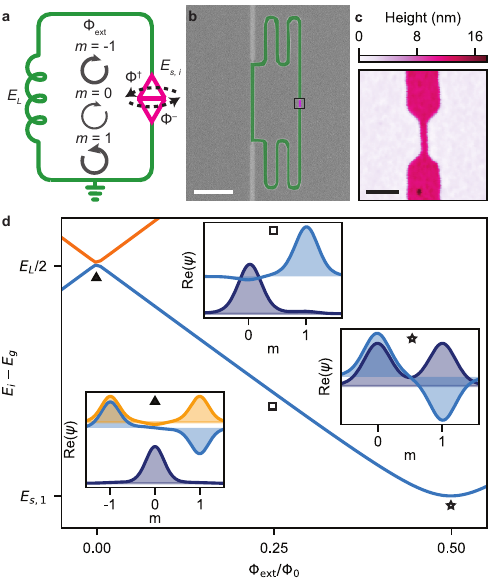}
\caption{\label{fig:circuit} \textbf{The phase-slip qubit.} (a) Circuit schematic of the device. The superinductance (green) is in parallel with a superconducting constriction (pink) where flux quanta can tunnel into and out of the loop at a rate $E_{s,\,i}$. 
The arrows indicate the strength and chirality of the current in the loop for fluxon states, $m$ = -1, 0, 1 when the external flux, $\Phi_\mathrm{ext}$, is slightly off of zero flux. 
(b) False-colored SEM image of lithographically similar device to Device A with the superinductance shown in green and the constriction shown in pink. 
The qubit is inductively coupled to a readout resonator (light gray).
Scale bar corresponds to \qty{5}{\mu m}.
(c) Atomic force microscopy image of the \qty{18}{nm} wide constriction forming the QPS junction.
Scale bar corresponds to \qty{100}{nm}.
(d) Transition energy spectrum from the ground state of a phase slip qubit as a function of \phiext.
At zero flux the qubit frequency is mainly determined by $E_L$, provided that $E_L\gg E_{s,i}$. 
The insets show the wavefunctions of the ground (navy), first (blue), and second excited (orange) states at \phiext=0, 0.25 and 0.5 $\Phi_0$.}
\end{figure}

In this work, we demonstrate the operation of a high-frequency superconducting qubit based on coherent quantum phase slips tunneling through a superconducting constriction. 
We fabricate the qubit using a single thin film of TiN. 
The qubit is biased to zero flux where the qubit basis states are similar to the recently-demonstrated bifluxon qubit \cite{kalashnikov_bifluxon_2020, ardati_using_2024, mencia_integer_2024} and the qubit frequency is mainly determined by the well-controlled qubit inductance. 
At this sweet spot, we demonstrate readout and coherent manipulation of the qubit. 
We observe qubit lifetimes in excess of \qty{60}{\mu s} and qubit coherence times of \qty{\sim17}{ns}. 
Finally, we exploit the higher energy gap of TiN to operate the qubit at temperatures \qty{>300}{mK}. 
The relaxation and coherence times of the qubit remain robust, with $T_{2R}$ unchanged and $T_1$ remaining above \qty{10}{\mu s}, at these temperatures. 

A phase slip qubit is formed from an inductance in parallel with a QPS junction, which creates a superconducting loop as shown in Fig.~\ref{fig:circuit}(a) \cite{mooij_phase-slip_2005}. 
The eigenstates in a continuous superconducting loop are persistent-current states that correspond to quantized flux quanta in the loop.
The addition of a QPS junction, {here implemented as a constriction in the loop} where quantum fluctuations of the superconducting order parameter are large, leads to flux quanta tunneling across the junction.
While the QPS junction is sometimes discussed as an independent circuit element \cite{astafiev_coherent_2012,le_doubly_2019}, we take here the point of view \cite{vanevic_quantum_2012} that a constriction will behave as a QPS junction only if shunted by a circuit with the right impedance.
If the constriction impedance dominates, then the current through the circuit will be determined by the tunneling rate of Cooper pairs across the constriction, resulting in the constriction acting as a JJ.
If the loop impedance significantly exceeds that of the constriction, then a small circulating current determined mainly by the loop will flow through the entire circuit and the constriction will act as a QPS junction. 
The resulting eigenstates of the circuit in the latter case are then coherent superpositions of different numbers of flux quanta in the loop.
The normal state resistance of the superconducting loop in our device is larger than that of the constriction by two orders of magnitude (See Supplementary Information), indicating that our constriction acts as a QPS junction.

The phase slip qubit is defined by the ground ($\ket{g}$) and first excited ($\ket{e}$) states of this device.
By tuning the flux through the phase slip circuit, \phiext, one can adjust the qubit energies and eigenstates as shown in Fig.~\ref{fig:circuit}(d).  
Qubits based on phase slips have been implemented with Al-based JJs, \cite{mooij_josephson_1999, manucharyan_fluxonium_2009, ardati_using_2024, mencia_integer_2024}, granular aluminum \cite{rieger_granular_2023}, and disordered superconductors \cite{astafiev_coherent_2012, peltonen_coherent_2016}. 
The main difference between these implementations are the energies at which the phase slip circuit shown in Fig.~\ref{fig:circuit}(a) no longer describes the device and it becomes necessary to include the contribution of parasitic capacitances.
Plasmon modes due to the capacitive contribution occur at a few GHz for circuits using Al-based JJs \cite{manucharyan_fluxonium_2009} but can be significantly higher in circuits using constrictions in disordered superconductors and granular aluminum \cite{astafiev_coherent_2012, rieger_granular_2023}. In our device, we estimate that the first plasma mode occurs above 50 GHz (see Supplementary Information).

At integer and half-multiple flux quantum, qubit eigenstates are equal superpositions of flux quanta through the loop and become insensitive to dephasing from fluctuations in the flux through the phase slip qubit. 
Previous implementations of constriction-based phase slip qubits have been operated at half-flux quantum \cite{astafiev_coherent_2012,peltonen_coherent_2016}. 
Half-flux quantum, however, is a difficult operating point for phase slip qubits made using constrictions due to the exponential sensitivity of the qubit energy on the width of the constriction. 

We focus instead on the integer-flux operating point of the phase slip qubit. 
Here, the qubit energy remains insensitive to flux noise while also being only weakly dependent on the phase slip rate through the QPS junction. 
The qubit frequency is primarily determined by $E_L$ in the limit $E_L \gg E_{s,\,i}$, where $E_L= \Phi_0^2 /L_k$ is the energy due to the inductance ($L_k)$ of the loop and 
$E_{s,\,i}$ is the rate of $i$ coherent quantum phase slips through the {Q}PS junction. 
The inductive energy is controlled reliably through the inductance of the phase slip qubit. 
$E_{s,\,i}$ determines the energy splitting between the first and second-excited ($\ket{f}$) states, which is a small perturbation on the qubit frequency.
At integer flux, the qubit is thus minimally sensitive to the geometric parameters of the constriction.
Moreover, the minimal overlap between the ground and excited state of the qubit eigenstates at integer flux provides increased protection against qubit relaxation \cite{kalashnikov_bifluxon_2020,ardati_using_2024,mencia_integer_2024}.
\section*{Results}
Our implementation of a phase slip qubit is shown in Fig.~\ref{fig:circuit}(b)-(c). 
The device is formed from a single layer of thin (\qty{5}{nm} thick) TiN, which was chosen due to its large kinetic inductance \cite{amin_loss_2022, joshi_strong_2022, shearrow_atomic_2018} and high measured quality factors \cite{amin_loss_2022, vissers_low_2010}. 
We fabricated the junction from a narrow constriction of width  \qty{\sim 18}{nm} (Fig.~\ref{fig:circuit}(c)), which is smaller than the expected coherence length of TiN to enable coherent quantum phase slips through the constriction \cite{saveskul_superconductivity_2019, faley_titanium_2021, bastiaans_direct_2021}.
Since uncontrolled phase slips across the {loop} inductance are a detrimental dephasing process for the phase slip qubit\cite{manucharyan_evidence_2012,randeria_dephasing_2024}, we designed the width of the inductance to be \qty{150}{nm}, which should be wide enough relative to the superconducting coherence length to prevent these spurious phase slips. 
The inductance is shown in green in Fig.~\ref{fig:circuit}(b).
We couple the phase slip device inductively to a readout resonator in order to perform dispersive readout using standard circuit QED techniques \cite{blais_quantum-information_2007}.
We have measured two devices (A and B), which showed similar results. We will discuss Device A in the main text while measurements from Device B can be found in the Supplementary Information.

        \begin{figure}
        \centering
            \includegraphics{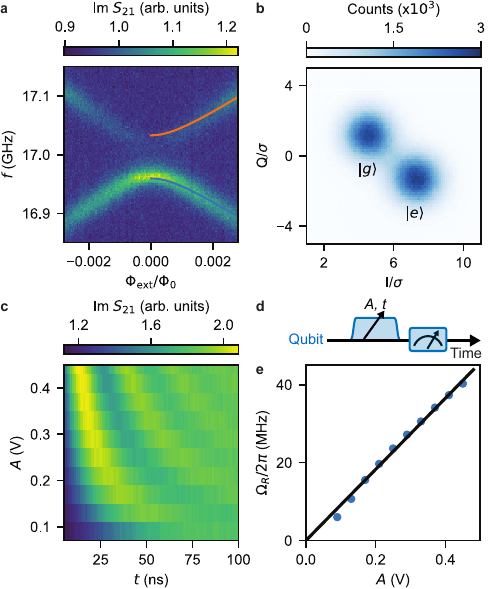}
            \caption{\label{fig:spec} \textbf{Qubit readout and control.}
                (a) Two tone spectroscopy near zero flux. The fit to the $\ket{g}$-$\ket{e}$ transition (blue line) and $\ket{g}$-$\ket{f}$ transition (orange line) is shown on the right half of the data.
                (b) Histogram of single-shot qubit readout. 
                The qubit is populated to both the ground and excited state via a $X_{\pi/2}$ pulse prior to measurement. 
                We readout the qubit at the resonator frequency corresponding to qubit being in the $\ket{g}$ state.
                (c) Rabi oscillations of the qubit versus a square pulse with room temperature drive amplitude $A$ and length $t$.
                (d) Pulse sequence for the Rabi experiment shown in (c).
                (e) Rabi frequency as a function of $A$. 
                We extract the Rabi frequencies from (c).
                The black line is a linear fit of the Rabi frequency versus drive amplitude.}
        \end{figure}

We first performed two-tone spectroscopy on the device to check that it is behaving as a phase slip qubit (Fig.~\ref{fig:spec}). We expect the behavior of the device to be described by the following Hamiltonian written in the fluxon basis:

\begin{equation} \label{eq1}
  \begin{split}
    H &= \frac{E_L}{2}\sum_{m}{[(m-\frac{\Phi_\mathrm{ext}}{\Phi_{0}}})^2\ket{m}\bra{m}] \\
      &\quad -\frac{E_{s,\, 1}}{2}\sum_{m}{(\ket{m}\bra{m +1}+\ket{m+1}\bra{m})} \\
&\quad + \frac{E_{s,\,2}}{2}\sum_{m}{(\ket{m}\bra{m+2}+\ket{m+2}\bra{m})}  \\
  \end{split}
\end{equation}

\noindent where $m$ is the number of fluxons in the loop, $E_{s,\,1}$ is the phase slip rate for coupling the fluxon states $m$ and $m +1$, and $E_{s,\,2}$ is the rate at which two flux quanta tunnel simultaneously, coupling the fluxon states $m$ and $m + 2$. 
We fit {both the $\ket{g}-\ket{e}$ and the $\ket{g}-\ket{f}$ transitions} to the Hamiltonian described by Eq.~\ref{eq1} (solid lines in Fig.~\ref{fig:spec}(a)) and observe good agreement between the data and theory. 
From the fit, we find $E_L =~$\qty{34.4}{GHz}, $E_{s,\, 1} =~$\qty{1.03}{GHz}, and $E_{s, \,2} =~$\qty{0.05}{GHz}. 
Equation~\ref{eq1} is an effective low-energy Hamiltonian that only holds for our circuit up to the energy of the first plasmon mode.
By fitting our data to the more general fluxonium Hamiltonian (fit parameters provided in Supplementary information), we find that the first plasmon mode is expected above 50 GHz, which indicates that Eq.~\ref{eq1} describes our experiments well.
As an additional check that the qubit is behaving as predicted, we note that this Hamiltonian enforces a parity selection rule at exactly zero flux for transitions between the $\ket{g}$ and $\ket{f}$ state, which we observe as a decrease in the visibility of that transition. 

We also demonstrate our ability to perform single-shot readout, which is essential in implementing future quantum algorithms and quantum error correction schemes. We drove the qubit at the $\ket{g}$-$\ket{e}$ frequency to populate the qubit in both the ground and excited states, and measured the resonator response.
We performed this measurement $2\times10^{6}$ times and constructed a histogram of the resulting data, which is presented in Fig.~\ref{fig:spec}(b). 
We observe clear separation between the $\ket{g}$ and $\ket{e}$ resonator responses with an assignment fidelity of 96\% \cite{gambetta_protocols_2007}.

        \begin{figure*}[t]
        \centering
            \includegraphics{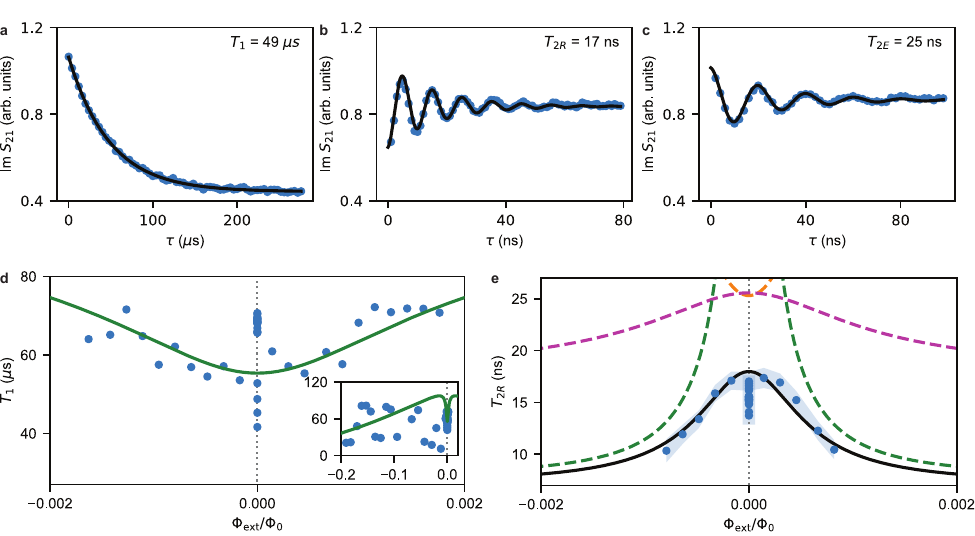}
            \caption{\label{fig:time} \textbf{Qubit relaxation and coherence.}
                (a-c) Time domain measurements taken at zero flux measuring (a) the relaxation time $T_1$, (b) the Ramsey coherence time $T_{2R}$, and (c) the coherence time with the application of a single echo pulse $T_{2E}$.
                The duration of the $X_{\pi}$ and $X_{\pi/2}$ pulses used in the measurements was \qty{9}{ns}.
                (d) Relaxation time $T_1$ versus flux. The expected flux dependence of $T_1$ due to inductive loss with $Q_\mathrm{ind} = 4.2 \times 10^4$ is plotted in green. The inset shows that $T_1$ deviates from the inductive loss fit away from zero flux. The gray dotted line is a guide for the eye for zero flux. The range of variations of $T_1$ in time is shown by the data points at zero flux.
                (e) Ramsey coherence time $T_{2R}$ versus flux. The shaded region indicates the error from fitting the $T_{2R}$ data. 
                The dashed lines indicate the individual contributions to dephasing of the qubit. The green line shows the expected flux dependence of $T_{2R}$ due to first-order flux noise dephasing with noise amplitude, $A_{\Phi} = 1.41\times 10^{-4} \Phi_0$/$\sqrt{\text{Hz}}$, 
                the pink line is the limit set by Aharonov-Casher dephasing for variation in $E_{s,\,1}$ with noise amplitude, $A_{\text{AC}} = 0.026E_{s,\,1}$/$\sqrt{\text{Hz}}$,
                the orange line indicates the limit set by thermal excitations to the $\ket{f}$ state for a qubit temperature of \qty{17}{mK}.
                The black line shows the expected limit from the quadrature sum of all three contributions.
                Similar to (d), the gray line is a guide for zero flux and the range of variations of $T_{2R}$ in time is shown by the data points at zero flux.
                Note that at zero flux, only the point with the highest $T_{2R}$ was included in the fit.
                }
        \end{figure*}

We next perform qubit control by driving Rabi oscillations between the ground state and the excited state. 
We applied a drive with amplitude, $A$, and varied the duration, $t$, before measuring the state of the qubit via a readout pulse as described by the sequence shown in Fig.~\ref{fig:spec}(d).
Figure~\ref{fig:spec}(c) shows oscillations in the qubit state as a function of $A$ and $t$. 
The extracted Rabi frequency for each pulse amplitude is plotted in Fig.~\ref{fig:spec}(e). 
The measured Rabi frequency depends linearly on the amplitude, which we expect for the behavior of a driven qubit. 
We note that despite the proximity of the $\ket{g}-\ket{f}$ transition to the $\ket{g}-\ket{e}$ transition, we do not observe any spurious oscillations in Fig.~\ref{fig:spec}(c). 
This further indicates that our phase slip qubit is behaving as a well-controlled two-level system.

We conducted time-domain measurements to characterize the qubit lifetime and coherence at the zero-flux sweet spot. 
We used standard measurement protocols to measure the relaxation time $T_1$, the Ramsey coherence time $T_{2R}$, and the single Hahn-echo coherence time $T_{2E}$ (Fig.~\ref{fig:time}(a-c)).
The observed $T_1$ improves upon previously-measured relaxation times in phase slip qubits at \phiext$=0.5\Phi_0$ by a factor of $10^3$.
The $T_{2R}$ is significantly smaller than $T_1$, however, which indicates strong qubit dephasing.
The coherence time only partly improves with an echo pulse, indicating that the qubit may be suffering from high-frequency noise. 
The echo may also have resulted in limited improvement because our $X_{\pi}$ pulse was imperfect due to our pulse width being comparable to $T_{2R}$.  

Next, we investigate the flux and time dependence of $T_1$ to identify the loss mechanisms affecting our phase slip qubit (Fig.~\ref{fig:time}(d)).
The relaxation time demonstrates well-defined flux dependence around zero flux, but fluctuates significantly as the flux is moved away from zero flux as shown in the inset of Fig.~\ref{fig:time}(d).
Near zero flux, we can model the flux dependence of the relaxation due to inductive loss \cite{smith_superconducting_2020} and find an inductive quality factor, $Q_\mathrm{ind} = 4.2 \times 10^4$.
{We speculate that the inductive loss originates from materials defects or excess quasiparticles in the superinductance \cite{smith_superconducting_2020}}.
The measured $T_1$ deviates significantly from the fit as we sweep away from zero $\Phi_0$.
From qubit spectroscopy in these regions, we suspect that that the qubit may be coupling to multiple two-level systems (TLS), which would unpredictably degrade the relaxation time and contribute to this deviation.
We have also measured $T_1$ over multiple days at zero flux and observed fluctuations from $T_1 =~40-70~\mu$s (shown in the multiple data points at zero flux in Fig.~\ref{fig:time}(d)).
We conjecture that this variability could also be due to spurious TLS coupling to the qubit, with the TLS landscape changing over time \cite{thorbeck_two-level-system_2023}.

Next we performed $T_{2R}$ measurements versus flux near zero flux to investigate the dephasing mechanisms of the qubit (Fig.~\ref{fig:time}(e)).
We consider three contributions to dephasing: flux noise \cite{koch_flicker_1983, ithier_decoherence_2005}, Aharonov-Casher (AC) noise \cite{manucharyan_evidence_2012, randeria_dephasing_2024}, and noise due to the proximity of the $\ket{f}$ state \cite{ardati_using_2024}, depicted by the green, pink and orange lines in Fig.~\ref{fig:time}(e), respectively.
Flux noise contributes strongly to the dephasing away from zero flux but the qubit becomes first-order insensitive to flux noise at zero flux. 
We thus expect the AC noise and thermal excitations to the $|f\rangle$ state to determine the coherence times at zero flux.
Possible sources for AC noise may arise from localized charge puddles or excess quasiparticles in the disordered superconductor \cite{de_graaf_dual_2020}.
The slow dephasing from AC noise has typically been mitigated using echo sequences \cite{randeria_dephasing_2024} but we do not observe a significant improvement in $T_{2E}$ over $T_{2R}$.
We therefore also consider fast dephasing due to the proximity of the $|f\rangle$ state.
Since the $\ket{f}$ state is located only $\sim$\qty{70}{MHz} away from the $\ket{e}$ state at zero flux, thermal excitation from $\ket{e}$ to $\ket{f}$ can cause dephasing \cite{ardati_using_2024}.
Fitting our data using a quadrature sum of all three mechanisms (black line in Fig.~\ref{fig:time}(e)), we find a flux noise amplitude of $A_{\Phi} = 1.41 \times 10^{-4}\Phi_0$/$\sqrt{\text{Hz}}$, an AC noise amplitude, $A_{\text{AC}}= 0.026E_{s,\,1}$/$\sqrt{\text{Hz}}$, and a qubit temperature of 17 mK (from dephasing to the $|f\rangle$ state). 
The extracted $A_{\Phi}$ is $\sim2$ orders of magnitude larger than current state-of-the-art Josephson junction-based flux qubits \cite{nguyen_high-coherence_2019}.
Impurities introduced during device fabrication may be responsible for this increased flux noise \cite{saveskul_superconductivity_2019}.
The low extracted qubit temperature is surprising when compared with our qubit histogram measurements and previous experiments on superconducting qubits \cite{bao_fluxonium_2022,rieger_granular_2023}.
We hypothesize that changing the geometry of the constriction will enable us to better identify the contributions associated with each dephasing mechanism and reduce the total dephasing experienced by the qubit.

        \begin{figure}
        \centering
            \includegraphics{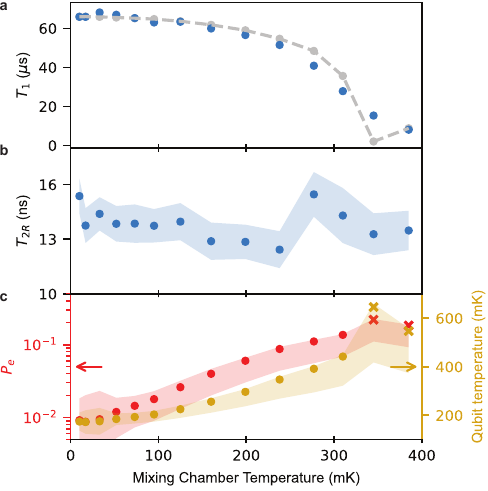}
            \caption{\label{fig:temp} \textbf{{Operation of the qubit at elevated temperatures.}}
                (a) Temperature dependence of qubit lifetime. The blue points are the measured $T_1$ values.
                The gray points with dashed lines represent the expected $T_1$ degradation due to decreasing $Q_\mathrm{ind}$ from the rise in the quasiparticle population as the qubit temperature increases. We use the measured qubit temperature as input to our model.
                (b) Temperature dependence of qubit coherence time.
                The shaded region is the error from the $T_{2R}$ fit.
                $T_{2R}$ does not show significant variation with temperature.
                (c) Excited state probability, $P_e$ (left) and qubit temperature (right) as a function of mixing chamber temperature.
                {The upper bound of the shaded region represent the uncertainty from the qubit histogram fit.
                The lower bound of the shaded region also accounts for the uncertainty associated with overestimating the $\ket{e}$ state population due to excess $\ket{f}$ state population in the qubit histogram.}
                The crosses represent data which was taken with a different readout parameter compared to the circles, which was chosen due to the reduction in $T_1$ at higher temperatures.}
        \end{figure}

Finally, we investigate the benefits of the higher qubit frequency and larger gap of TiN in operating the phase slip qubit at higher temperatures. 
Figure~\ref{fig:temp}(a-b) shows the measured relaxation time and coherence time of the qubit up to 385 mK. 
We observe that $T_1$ remains above \qty{20}{\mu s} for temperatures up to \qty{300}{mK} while $T_{2R}$ remains essentially constant with temperature. 
The relaxation time shows a slow degradation up to \qty{240}{mK}, after which $T_1$ decreases significantly.
Previous measurements of aluminum-based superconducting qubits have also observed a significant decrease in $T_1$ at temperatures around 150 mK, which was attributed to an increase in quasiparticle loss \cite{paik_observation_2011, serniak_hot_2018,martinis_energy_2009}. 
More recent experiments have demonstrated that the operating temperature of Al JJ-based qubits could be increased by proximitizing the junction using Nb, but observed short lifetimes of about \qty{1}{\mu s} for qubits operating at higher frequencies $>10$~GHz \cite{anferov_superconducting_2024}.

The decreased $T_1$ in our work is unexpected since the measured transition temperature of our TiN film is \qty{2.9}{K}. 
We additionally measured the qubit excited-state population as a function of temperature to better understand the temperature dependence of the relaxation time as shown in Fig.~\ref{fig:temp}(c).
{In the qubit histograms for these measurements, we observe two separated resonator responses corresponding to the $\ket{g}$ and $\ket{e}$ states.}
The measured probability {from these histograms} is converted to qubit temperature using the Boltzmann distribution (right axis in Fig.~\ref{fig:temp}(c)).
The qubit temperature appears to be consistently greater than the mixing chamber temperature by \qty{>100}{mK}.
{We note that due to the fast decay rate of $\ket{f}$ to $\ket{e}$, we may be overestimating the $\ket{e}$ population in our readout histograms due to thermal population of $\ket{f}$.
This possible overestimation is reflected in the lower bound of the error bars in Fig.~\ref{fig:temp}(c) for the qubit temperature and population data.}
Using the measured qubit temperature, we model the reduction in lifetime as inductive loss, with $Q_\mathrm{ind}$ changing as a function of qubit temperature to reflect an increasing quasiparticle density as shown by the gray points with dashed lines in Fig.~\ref{fig:temp}(a) (See Supplementary Information). 
We find reasonable agreement of the measured relaxation times to this model.

\section*{Discussion}
We have demonstrated all essential single-qubit functions in a phase slip qubit operated at zero flux fabricated from a single layer of TiN. 
Our results show that a superconducting constriction is a viable alternative source of nonlinearity that can be used in coherent quantum devices. 
Having multiple superconducting sources of nonlinearity significantly expands the possible quantum coherent devices that can be built.
First, since the QPS junction effectively behaves as a nonlinear capacitor rather than a nonlinear inductor (up to an expected frequency of 53 GHz in our case), it would allow for the creation of novel oscillators with photon-dependent non-linearities \cite{hriscu_model_2011} that can be used for sensitive detection and signal amplification.
Second, TiN has been demonstrated to have higher quality factors than evaporated aluminum \cite{amin_loss_2022}. 
Our demonstrated phase-slip qubit is thus a first step toward qubits with longer possible relaxation times than current AlOx-junction based superconducting qubits.
Achieving longer qubit coherence times will be necessary for the {constriction-based} phase slip qubit to become a competitive platform for quantum information processing. 
While our qubit has relaxation times comparable to standard JJ-based superconducting qubits, the coherence times of the qubit are very short.
Improved cleaning methods in the device processing or changes in the geometry to reduce {the sensitivity of phase slip tunneling processes to fluctuating charges} will further elucidate these decoherence mechanisms.

Our phase slip qubit also presents a platform for building high frequency qubits that can operate at high temperatures.
Building superconducting qubits that can operate at high temperatures has become increasingly important as superconducting quantum processors run into limits imposed by cooling capabilities of current technologies. 
While we observed that the qubit relaxation time started to degrade significantly at temperatures below the superconducting transition temperature, our demonstrated qubit operation above \qty{100}{mK} already significantly reduces the technological requirements needed to cool superconducting qubits beyond what can be achieved by aluminum-based superconducting qubits. 
The observed poor thermalization of the qubit may be due to it coupling to a different bath compared with current superconducting qubits, which would require additional filtering compared to standard devices. 
The very thin superconducting TiN film may also have large enough disorder to prolong thermalization times; measurements of qubits made using superconducting films with different levels of disorder will be useful in understanding this mechanism.
Usage of superconducting films with even higher transition temperatures than TiN may present a path to superconducting qubits operating above \qty{4}{K}.

\section*{Methods}
\subsection{Device fabrication}
We fabricated the sample from a \qty{5}{nm} thick TiN film on a \qty{720}{\mu m} thick silicon substrate. 
Smaller features such as the resonator and the qubit were fabricated using electron-beam lithography (Elionix ELS-G150) with ZEP 520A resist. 
We used photolitography (Heidelberg MLA 150) with AZ1518 resist to pattern larger features such as the ground plane and transmission line. 
The sample was etched after each patterning step using an Oxford Mixed ICP-RIE system with a chemistry of $\text{Cl}_2$, $\text{BCl}_3$, and $\text{Ar}$.
To clean the sample, we used an $\text{O}_2$ descum process, followed by rinsing in N-Methylpyrrolidone (NMP). 
Prior to wirebonding to a printed circuit board, the sample was rinsed in buffered oxide echant (BOE) to remove surface oxides.
\subsection{Experimental Setup}
The sample was enclosed in a Copper sample holder and cooled to 10 mK in a dilution refrigerator (Oxford Instruments, Triton 500).
 A coil was mounted to the sample holder and a DC current source (Yokogawa, GS200) was used to apply magnetic field to the sample.
The readout and qubit pulses were formed by mixing local oscillator(LO) tones from signal generators with IQ tones from an integrated FPGA system (Quantum Machines, OPX +).
The input signal was thermalized via attenuators at different temperature plates before reaching the mixing chamber.
At the mixing chamber plate, the input was sent through an Eccosorb filter to filter out high-frequency noise before entering the transmission line.

The output from the transmission line was routed through another Eccosorb filter followed by a traveling wave parametric amplifier (TWPA). The output signal was then amplified again at the 4K plate via a high-electron-mobility transistor (HEMT) amplifier (Low Noise Factory, $\text{LNF-LNC4\_16B}$) and at room temperature via a low noise amplifier ($\text{LNF-LNR4\_14B\_SV}$). The signal was demodulated and further amplified before being sent to the FPGA for processing. 
\subsection{Simulations}
Prior to measurements, we estimated the resonator frequency and coupling quality factor using electromagnetic field simulation (Ansys HFSS). To solve the Hamiltonian and estimate qubit frequency, we employed numerical diagonalization in QuTip \cite{johansson_qutip_2013}. We also used QuTip to fit the qubit spectroscopy and perform master equation pulse-level simulations (See Supplementary Information).

\section*{Data availability}
The experimental data that support the findings of this study is available in the Illinois Data Bank at \url{https://doi.org/10.13012/B2IDB-0982994_V1} (ref. \cite{purmessur_2025})

\section*{Code availability}
The code used for the data analysis and visualization is available at is available in the Illinois Data Bank at \url{https://doi.org/10.13012/B2IDB-0982994_V1} (ref. \cite{purmessur_2025})

\bibliography{maintext1}

\section*{Acknowledgments}
We thank Matthias Steffen, Benjamin Wymore, Oliver Dial, Aayam Bista, Xi Cao, Jared Gibson, Jinwoong Philip Kim, Michael Mollenhauer, Ke Nie, and Randy Owen for useful discussions. The TiN film and traveling-wave parametric amplifier used in this experiment were provided by IBM. This research was carried out in part in the Materials Research Lab Central Facilities and the Holonyak Micro and Nanotechnology Lab, University of Illinois. C.P. and A.K. are funded through the IBM-Illinois Discovery Accelerator Institute.

\section*{Author Contributions}
C.P. designed and measured the devices. K.C. and C.P. fabricated the devices. C.P. performed the data analysis with input from B.v.H. and A.K. C.P. and A.K. wrote the manuscript, with feedback from all authors. A.K. conceived and supervised the experiment.

\section*{Competing Interests}
The authors declare no competing interests.

\nocite{*}

\end{document}


\title{Supplementary Information for Operation of a high-frequency,  phase-slip qubit}

\author{Cheeranjeev Purmessur}
\affiliation{Department of Physics, University of Illinois Urbana-Champaign, Urbana, IL 61801}

\author{Kaicheung Chow}
\affiliation{Holonyak Micro $\&$ Nanotechnology Lab, University of Illinois Urbana-Champaign, Urbana, IL 61801}
\author{Bernard van Heck}
\affiliation{Dipartimento di Fisica, Sapienza Università di Roma, Piazzale Aldo Moro 2, I-00185 Rome, Italy}
\author{Angela Kou}\thanks{Corresponding author: akou@illinois.edu}  
\affiliation{%
Department of Physics, University of Illinois Urbana-Champaign, Urbana, IL 61801}
\affiliation{Holonyak Micro $\&$ Nanotechnology Lab, University of Illinois Urbana-Champaign, Urbana, IL 61801}
\affiliation{Materials Research Laboratory,
University of Illinois at Urbana-Champaign, Urbana, IL 61801}
\date{\today}

\maketitle

\section{Full device}
\begin{figure*}[ht]
\centering
    \includegraphics{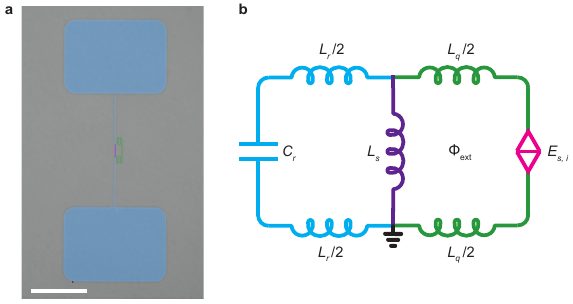}
    \caption{\label{fig:fulldevice} \textbf{Full device}
    (a) False-colored SEM image of a lithographically similar device to device A showing the resonator and qubit. 
    The scale bar corresponds to \qty{50}{\mu m}.
    (b) Circuit schematic of the full device.
    In (a) and (b), the qubit (green) is inductively coupled to the resonator (blue) via a shared inductance (purple). The PSJ is shown in pink. 
    }
\end{figure*}

The full device including both the lumped-element resonator and the qubit is shown in Fig.~\ref{fig:fulldevice}. 
The lumped-element resonator consists of two capacitor pads and an inductance formed from a narrow strip (width \qty{570}{nm}) of TiN.
The ground plane (not shown in the figure) is located \qty{15}{\mu m} from the top and bottom edges of pads and \qty{125}{\mu m} from the left and right edges.
The ground plane contains holes to trap magnetic vortices present around the device. 
Input readout and qubit pulses were coupled to the device via a transmission line. 
From resonator spectroscopy at zero flux, we obtain a coupling quality factor, $Q_c=1.8\times 10^4$ and an internal quality factor, $Q_i=4.9\times10^5$ when driving with a single photon.

The full Hamiltonian of the qubit coupled to the resonator is given by
\begin{equation} \label{totalH}
H_\mathrm{sys} = H_{q} +H_{r} +H_{c}
\end{equation}
where $H_{c}$ is the interaction Hamiltonian, representing the coupling between the resonator and the qubit. 
Below we write the resonator $H_r$, qubit $H_q$, and interaction Hamiltonians (with $h = 1$).

\begin{equation} \label{Hq}
    H_{q} = \frac{E_L}{2}\sum_{m}{[(m-\frac{\Phi_\mathrm{ext}}{\Phi_{0}}})^2\ket{m}\bra{m}] -\frac{E_{s,\, 1}}{2}\sum_{m}{(\ket{m}\bra{m +1}+\ket{m+1}\bra{m})} + \frac{E_{s,\,2}}{2}\sum_{m}{(\ket{m}\bra{m+2}+\ket{m+2}\bra{m})} 
\end{equation}

\begin{equation} \label{Hres}
    H_{r} = 4E_{Cr}\hat{n}_{r}^2+\frac{1}{2}E_{Lr}\hat{\phi}_r^2
\end{equation}

\begin{equation} \label{Hint}
    H_{c} = g\sum_{m}{(m-\frac{\Phi_\mathrm{ext}}{\Phi_{0}}})\ket{m}\bra{m}\times\frac{\hat{\phi}_r}{\sqrt{\zeta/2}}
\end{equation}

\noindent where $\hat{n}_r$ and $\hat{\phi}_r$ are the Cooper pair number and phase operators of the resonator with $\hat{n}_r=(1/\sqrt{2\zeta})i(a-a^\dagger)$ and $\hat{\phi}_r = \sqrt{\zeta/2}(a+a^\dagger)$.
Here, $\zeta = \sqrt{8E_{Cr}/E_{Lr}}$ is defined as the resonator impedance and $a$ and $a^{\dagger}$ are the ladder operators.
$E_{Cr}$, $E_{Lr}$, and $g$ are the charging energy, inductive energy, and coupling strength, respectively. 
To express $E_{Cr}$, $E_{Lr}$, $g$, and $E_L$ in terms of $L_r$, $C_r$, $L_s$, and $L_q$, we applied Kirchhoff's laws and solved for the Hamiltonian of the circuit (Fig.~\ref{fig:fulldevice} (b)).
We define the following terms:

\begin{equation} \label{Lt}
L_T' = L_qL_r +L_sL_r+L_sL_q\\
\end{equation}
\begin{equation} \label{Lq}
\frac{1}{L_q'} = \frac{1}{L_T'}(L_s+L_r)\\
\end{equation}
\begin{equation} \label{Lr}
\frac{1}{L_r'} = \frac{1}{L_T'}(L_s+L_q)\\
\end{equation}
Here we refer to $L_q'$ and $L_r'$ as the effective qubit and resonator inductance, respectively. 
Then,

\begin{equation} \label{Ecr}
E_{Cr} = \frac{e^2}{2C_r},~E_{Lr} = \left(\frac{\Phi_0}{2\pi}\right)^2\frac{1}{L_r'},~g = \frac{\Phi_0}{2\pi}\Phi_0\frac{L_s}{L_t'}\sqrt{\frac{\zeta}{2}},~E_L = \frac{\Phi_0^2}{L_q'}\\
\end{equation}

\section{Full qubit spectrum and resonator spectrum fit}
\begin{figure*}[ht]
\centering
    \includegraphics{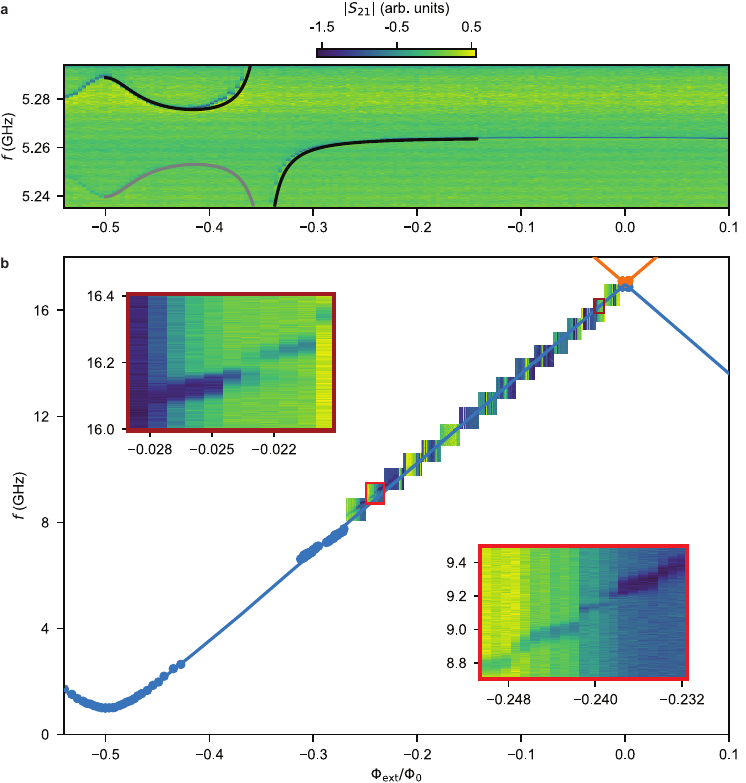}
    \caption{\label{fig:fullspec} \textbf{Resonator spectrum and full qubit spectrum}
    (a) Resonator spectroscopy fit and data. 
    The black line denotes the $\ket{g,0}-\ket{g,1}$ transition while the gray line denotes the $\ket{e,0}-\ket{e,1}$ transition.  
    From the fit, $f_r =$ \qty{5.264}{GHz} and $g=$ \qty{1.146}{GHz}.
    (b) Qubit spectroscopy fit and data. The blue and orange points are the $\ket{g}-\ket{e}$ and $\ket{g}-\ket{f}$ frequency points, respectively, that are used for fitting. 
    The flux of these points have been calibrated using the resonator frequency points in those regions. 
    The 2D color maps on the slope of the $\ket{g}-\ket{e}$ shows the coupling of the qubit to multiple TLS's. 
    }
\end{figure*}

In Fig.~\ref{fig:fullspec}, we show the resonator and qubit spectroscopy data for $\Phi_\mathrm{ext}/\Phi_\mathrm{0}  \approx[-0.56, 0]$ .
We fitted both sets of data to our system Hamiltonian and obtained $f_r =$ \qty{5.264}{GHz} and $g=$ \qty{1.146}{GHz} along with the $E_L$, $E_{s,\,1}$, and $E_{s,\,2}$ values mentioned in the main text.
From the fitted $E_L$ value, we estimate a kinetic inductance per square, $L_k/\square=380$ pH/$\square$.
For the qubit spectroscopy fit, we used all the blue data points for the $\ket{g}-\ket{e}$ transition and the orange data points for the $\ket{g}-\ket{f}$ transition in Fig.~\ref{fig:fullspec}(b). 
We calibrated the flux to applied current conversion using qubit measurements close to the sweet spots $\Phi_\mathrm{ext}=0, 0.5 \Phi_0$, and measurements of the resonator frequency versus flux.
In the region between $\Phi_\mathrm{ext}=-0.31 \Phi_0$ and $\Phi_\mathrm{ext}=-0.01 \Phi_0$, we observe flux hysteresis as well as a large number of two-level systems (TLS) coupled to the qubit, which altered the flux dependence of the qubit spectrum.
For the region between $\Phi_\mathrm{ext}=-0.31 \Phi_0$ and $\Phi_\mathrm{ext}=-0.27 \Phi_0$, the resonator frequency has a strong flux dependence due to the dispersive shift induced by the qubit, which allowed us to independently identify the flux location of qubit spectroscopy points in the presence of flux jumps and coupled TLS modes. 
At points $\Phi_\mathrm{ext}=[-0.27,-0.01] \Phi_0$, however, the resonator spectrum becomes flat versus flux, which made precise calibration of the flux difficult in this region.
These data points are shown as color maps with our best estimate of the flux and are not included in the fit (Fig.~\ref{fig:fullspec}(b)). 

\section{Inclusion of the \texorpdfstring{$E_{s,\,2}$}{Es} term}

\begin{figure*}[ht]
\centering
    \includegraphics{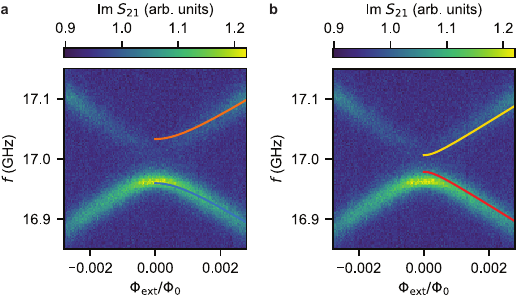}
    \caption{\label{fig:twotone_allfits} \textbf{Fitting to $H_\mathrm{sys}$ with and without the $E_{s,\,2}$ term}
    (a) Fit to $H_\mathrm{sys}$ including both the $E_{s,\,1}$ and $E_{s,\,2}$ terms. 
    The blue line is the $\ket{g}-\ket{e}$ fit and the orange line  is $\ket{g}-\ket{f}$ fit.
    (b) Fit to $H_\mathrm{sys}$ excluding the $E_{s,\,2}$ term. 
    The red line is the $\ket{g}-\ket{e}$ fit and the yellow line  is $\ket{g}-\ket{f}$ fit.
    }
\end{figure*}

In $H_\mathrm{sys}$, the $E_{s,\, 1}$ term determines the $\ket{g}$ to $\ket{e}$ energy splitting at half flux.
The $E_{s,\,2}$ term does not contribute to this splitting because, at half flux, the states involved differ by only one fluxon. 
Our data at half flux thus fixes the value of $E_{s,\, 1}$.
Near zero flux, transitions from the ground state to the $\ket{e}$ and $\ket{f}$ states involve tunneling of two fluxons, which can be sequential or simultaneous. 
Thus it is possible that $E_{s,\,2}$, which describes simultaneous flux tunneling, contributes to the energy splitting.
In Fig.~\ref{fig:twotone_allfits}, we fit the two tone spectroscopy near zero flux to  (a) $H_\mathrm{sys}$  and (b) $H_\mathrm{sys}$ without the $E_{s,\,2}$ term. 
The fitted parameters extracted for $H_\mathrm{sys}$ without  $E_{s,\,2}$  are $E_L$ = \qty{34.4}{GHz} and $E_{s,\,1}$ = \qty{1.03}{GHz}.
Clearly in (b), if we exclude  $E_{s,\,2}$, the fit underestimates the energy splitting between the  $\ket{e}$ and $\ket{f}$ state. Therefore, $E_{s,\, 2}$ is necessary to describe the measured $\ket{e}$ and $\ket{f}$ energy splitting.

\section{Comparison between phase slip Hamiltonian and fluxonium Hamiltonian}

\begin{figure*}[ht]
\centering
    \includegraphics{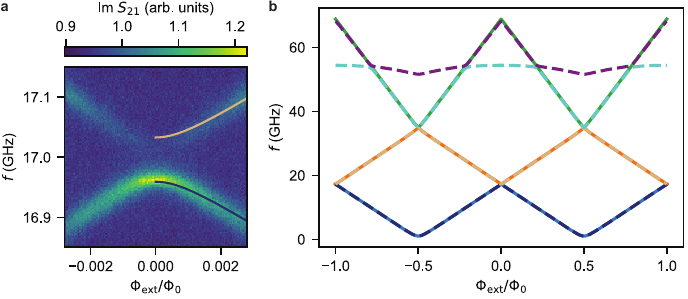}
    \caption{\label{fig:twotone_PSfluxfits} 
    \textbf{Fluxonium Hamiltonian fit and comparison of higher energy levels}
    (a) Fluxonium Hamiltonian fit at zero flux. The navy blue line is the $\ket{g}-\ket{e}$ transition and the beige line is the $\ket{g}-\ket{f}$ transition
    (b) Comparison of higher energy level transition between the phase slip Hamiltonian and the fluxonium Hamiltonian. The solid lines represent the phase slip qubit transitions while the dashed lines represent the fluxonium qubit transitions. The spectra deviates at $\sim$ \qty{53}{GHz}, where we observe a plasmon mode from the fluxonium spectrum that is not present in the ideal phase slip spectrum.
    }
\end{figure*}

As described in the main text, our device obeys the phase slip Hamiltonian and provides an impedance environment which favors flux tunneling across the constriction.
However, it is possible that because of parasitic capacitances, our qubit spectrum may deviate from the ideal phase slip Hamiltonian at higher energy levels.
Here we fit the data to the fluxonium Hamiltonian \cite{manucharyan_fluxonium_2009} below to roughly estimate where this deviation may occur.

\begin{equation} \label{eqfluxonium}
H_f = 4E_{Cf}\hat{n}^2_f+\frac{1}{2}E_{Lf}\hat{\phi}_f^2-E_J\cos{(\hat{\phi}_f+\phi_{ext})}
\end{equation}

\noindent where $E_{Cf}$ is the charging energy, $E_{Lf}$ is the inductive energy, and $E_J$ is the Josephson energy. $\hat{n}$ and $\hat{\phi}$ are the dimensionless charge and flux operators associated with the capacitor and the inductor in the fluxonium circuit. We expect the qubit spectrum to also fit the fluxonium Hamiltonian well since it is  more general than the phase slip Hamiltonian, accounting for contributions from any capacitance and inductance in the circuit. The fit around the zero flux point is shown in Fig.~\ref{fig:twotone_PSfluxfits}(a).
For consistency, we also include the interaction between the resonator and the fluxonium in the fit.
We extract $E_J = 50.0\text{ GHz}$, $E_{Cf} = 12.7\text{ GHz}$ and $E_{Lf} = 0.90\text{ GHz}$.
Fig.~\ref{fig:twotone_PSfluxfits}(b) shows the  higher energy levels from the phase slip Hamiltonian and the fluxonium Hamiltonian using the fitted parameters.
Transitions from both Hamiltonians are effectively identical up to the $\sim$\qty{53}{GHz} plasmon mode associated with the fluxonium Hamiltonian.
Therefore, we expect the phase slip Hamiltonian to describe our circuit well until at least $\sim$\qty{53}{GHz}.

Given that both fits are effectively identical for the lower-energy transitions, we can map the fluxonium Hamiltonian parameters to the phase slip Hamiltonian parameters~\cite{mencia_integer_2024, randeria_dephasing_2024, ardati_using_2024}.
In \cite{mencia_integer_2024}, $E_L$ is obtained by considering the total inductance in the fluxonium circuit and in \cite{randeria_dephasing_2024, ardati_using_2024}, $E_{s, 1}$ is obtained using the WKB approximation.
The expressions relating the parameters are shown below.

\begin{equation} \label{eqElEs1fromflux}
E_{L}= 4\pi^2\left(\frac{1}{E_{Lf}}+\frac{1}{E_{J}}\right)^{-1},~E_{s,1}= 2\sqrt{\frac{2}{\pi}}\sqrt{8E_JE_{Cf}}\sqrt[4]{\frac{8E_J}{E_{Cf}}}\exp{\left ( -\sqrt{\frac{8E_J}{E_{Cf}}}\right)}
\end{equation}

Plugging in our fitted fluxonium Hamiltonian parameters to the equations in Eq.~\ref{eqElEs1fromflux}, we obtain an estimated inductive
energy, $E_L=$ \qty{34.8}{GHz} and phase slip rate, $E_{s, 1}= $ \qty{0.99}{GHz}, which matches well with the values obtained from the
experimental fit.
We do not have an analytical expression using fluxonium parameters for $E_{s,2}$.
From perturbation theory, the second-order contribution from $E_{s,1}$ is given by $\sim E_{s,1}^2/E_L$ which is \qty{0.03}{GHz}.
This underestimates the measured $\ket{e}-\ket{f}$ splitting, which justifies the use of the $E_{s,2}$ term.

\section{Estimating sheet inductance and phase slip rate from film parameters}

Given the normal state sheet resistance, $R_{sq}$, and the $T_c$ of our \qty{5}{nm} TiN film, we can estimate the sheet inductance, $L_k/\square$, and the phase slip rate, $E_{s, 1}$ of our device.
Using $L_k/\square = \hbar R_{sq}/\pi\Delta_0$ \cite{shearrow_atomic_2018}, where $\Delta_0 \approx 1.76k_BT_c$ is the superconducting gap of TiN at zero temperature, we obtain $L_k/\square =340 \text{ pH}/\square$.
This is only around \qty{10}{\%} off from the $380 \text{ pH}/\square$ value extracted using the qubit spectrum fit.
Note that here $R_{sq} = 720\space\Omega/\square$ is extracted from room-temperature resistance measurements.
The transition temperature $T_c = 2.9\text{ K}$ is obtained from independent measurements of the resonator frequency with respect to temperature \cite{joshi_strong_2022}.

We can follow the calculations in \cite{vanevic_quantum_2012} to estimate the phase-slip rate across the constriction and across the loop from measured normal-state resistances.
First, we divide the wire from the loop and the constriction into sections of length, $l_c = 3.43\space\xi$, where  $\xi$ is the coherence length. 
The coherence length can range from  \qty{10}{nm} to \qty{105}{nm} according to literature \cite{bastiaans_direct_2021,saveskul_superconductivity_2019,faley_titanium_2021}.
The phase slip rate of one section of a wire is given by: 

\begin{equation} \label{psestimate}
E_{s}^*=1.08\Delta_0\sqrt{\frac{R_Q}{R_{\xi}}}\exp(-0.36\frac{R_Q}{R_{\xi}}),
\end{equation}

\noindent where $R_{Q}=h/2e^2$ and $R_{\xi}$ is the normal state resistance of a wire of length $\xi$ \cite{vanevic_quantum_2012}.
Given a loop length of $L_{\text{loop}}=$ \qty{65}{\mu m} and a width of \qty{150}{nm}, we obtain a range for $E_s^*$ from $1.6\times10^{-41}\Delta_0-5.4\times10^{-4}\Delta_0$ (for $\xi=10-105 \text{ nm}$).
For the constriction of length, $L_{\text{cons}}=90\text{ nm}$ and width \qty{18}{nm}, we get $E_{s}^*=5.5\times10^{-5}\Delta_0-0.63\Delta_0$.
To obtain $E_{s}^{\text{loop}}$ and $E_{s}^{\text{cons}}$, we multiply $E_{s}^*$ by the number of sections; $E_s = E_{s}^*L/l_c$. We then obtain $E_{s}^{\text{loop}} = [3.2\times10^{-36},10.3]\text{ GHz}$ and $E_{s}^{\text{cons}} = [0.015,16.6]\text{ GHz}$.
Our measured $E_{s,1}$ falls within the range for $E_{s}^{\text{cons}}$.
We note that for the calculation and the measured $E_{s,1}$ to coincide, $\xi$ should be set to \qty{18}{nm}.
Since the phase slip rate depends exponentially on $R_{\xi}$, a small overestimation of $\xi$ or underestimation of the film resistance at the constriction can lead to a large change in the calculated value.

From our parameters, we also find that the normal state resistance of the inductive loop is \qty{312}{k\Omega}, while that of the constriction is \qty{3.6}{k\Omega}.
This establishes an impedance environment where  $R_{\text{loop}}\gg R_{\text{cons}}$. 
Therefore, the device is in a regime where phase slips are localized to the constriction site and where the constriction  effectively acts as a phase slip junction \cite{vanevic_quantum_2012}.
\section{Master Equation Simulation}
\begin{figure*}[ht]
\centering
    \includegraphics{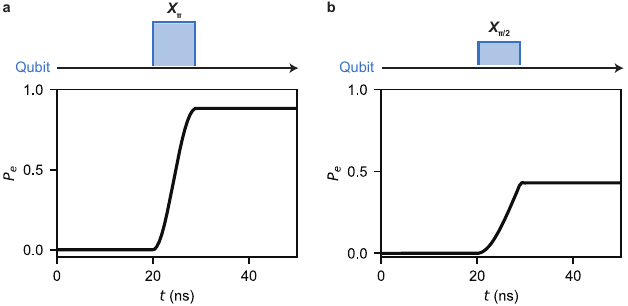}
    \caption{\label{fig:mastereq} \textbf{Pulse-level master equation simulation results} (a) and (b) show the probability of the qubit being in the $\ket{e}$ state after applying a $X_{\pi}$ and $X_{\pi/2}$ pulse, respectively. The duration of the pulses is set to \qty{9}{ns}, which is the pulse length used in our experiment.
    }
\end{figure*}

Given that our measurement pulse duration (\qty{9}{ns}) is not significantly shorter than the qubit coherence time, we suspect that our $X_{\pi}$ and $X_{\pi/2}$ measurement pulses are imperfect. 
They would therefore not bring the qubit completely to the $\ket{e}$ or $\frac{1}{2}(\ket{g}+\ket{e})$ states during the time domain measurements. 
More specifically, for the qubit relaxation measurement, the qubit is most likely not $100\%$ in the $\ket{e}$ state to start with but rather in some mixture of $\ket{g}$ and $\ket{e}$. 
We conduct a pulse-level master equation simulation in QuTip to quantify the imperfection of our $X_{\pi}$ and $X_{\pi/2}$ pulses \cite{johansson_qutip_2013}.
In the simulation, we specify the relaxation and dephasing rates at zero flux in the collapse operators. 
The results of these simulations can be found in Fig.~\ref{fig:mastereq}. 
We obtain a probability of being in the excited state, $P_e=0.88$ after applying a $X_{\pi}$ pulse and $P_e=0.43$ after applying a $X_{\pi/2}$ pulse.

\section{Qubit lifetime vs flux analysis}
\begin{figure*}[ht]
\centering
    \includegraphics{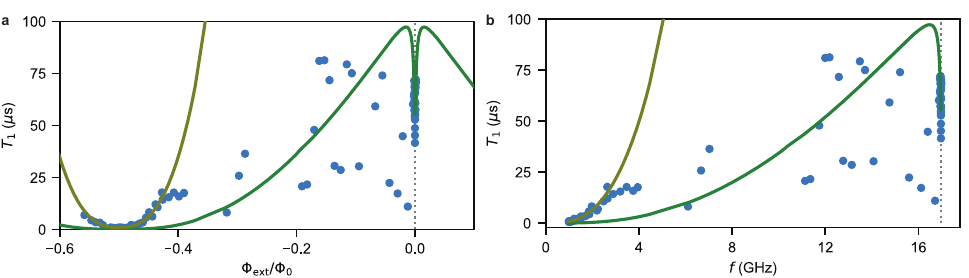}
    \caption{\label{fig:t1_indloss} \textbf{Full qubit lifetime data and fit }
    (a) $T_1$ as a function of flux. 
    (b) $T_1$ as a function of qubit frequency. 
    In (a) and (b), the expected $T_1$ due to inductive loss with $Q_\mathrm{ind} = 6.6\times10^5$ is plotted in olive green and agrees well with the $T_1$ data taken around  $-0.5\Phi_\mathrm{0}$.
    The darker green line is the fit around $0\Phi_\mathrm{0}$ with $Q_\mathrm{ind} = 4.2\times10^ 4$. 
    The data in the range $\Phi_\mathrm{ext}/\Phi_\mathrm{0}  \approx[-0.4, -0.01]$ or $f  \approx[4, 16.8]$ GHz deviates significantly from the fits.
    We attribute this deviation to the presence of TLS residing at specific qubit frequencies.
    }
\end{figure*}

We model the flux dependence of the relaxation time using the inductive loss model described in \cite{smith_superconducting_2020}. 
Employing Fermi's golden rule, we can relate the relaxation rate due to inductive loss to the matrix element of the flux across the superinductance, $\bra e\hat\Phi\ket g$, and the power spectral density of the noise, $S(\omega_{ge})$.

\begin{equation} \label{eqinductiveloss}
\frac{1}{T_{1}}=\Gamma_{1} = \frac{1}{\hbar^2}\left|\bra e\hat\Phi\ket g\right| ^2 [S(\omega_{ge})+S(-\omega_{ge})]
\end{equation}
In the expression above, the relaxation rate due to both emission and absorption is included. 
The sum of the noise spectral densities is given by

\begin{equation} \label{eq_noisespectal}
S(\omega_{ge})+S(-\omega_{ge}) = \frac{2\hbar}{LQ_\mathrm{ind}}\text{coth}\frac{\hbar|\omega_{ge}|}{2k_BT}
\end{equation}
Here $L$ is the effective qubit inductance, $Q_\mathrm{ind}$ is the inductive quality factor and $T$ is the bath temperature. 
We assume that  $T=$ \qty{174}{mK}, which is the temperature of the qubit at the base mixing chamber temperature (See Fig. 4 of the main text). 
In Fig.~\ref{fig:t1_indloss} (a), we show the $T_1$ data that was taken in the range $\Phi_\mathrm{ext}/\Phi_\mathrm{0}  \approx[-0.56, 0]$. 
We see a clear flux dependence for data around $-0.5\Phi_\mathrm{0}$.
The olive green line is the fit for the data in this region, where we obtain $Q_\mathrm{ind} = 6.6\times 10^5$.
The darker green line is the expected $T_1$ values when $Q_\mathrm{ind} = 4.2\times10^4$.
This line is also shown near zero flux in Fig. 3(d) of the main text.
It appears that the $Q_\mathrm{ind}$ varies at different flux points (or qubit frequencies). 
From \cite{barends_minimizing_2011}, we can relate the quality factor to the quasiparticle density ratio, $x_{qp}$:

\begin{equation} \label{eq_noisespectal}
\frac{1}{Q}= \frac{\alpha}{\pi}\sqrt{\frac{2\Delta}{hf}}x_{qp}
\end{equation}
\noindent In the expression above, $\alpha = L_k/L_\mathrm{total}$ is the kinetic inductance fraction, which is approximately one in our devices. At zero flux when $Q_\mathrm{ind} = 4.2\times10^4$, $x_{qp}=2.1\times10^{-5}$, while at half flux when $Q_\mathrm{ind} = 6.6\times 10^5$, $x_{qp}=3.3\times10^{-7}$. 
The quasiparticle population is higher when measuring at higher frequencies. 
This increase may be due to the device being driven by frequencies in closer proximity to the superconducting gap.
Fig.~\ref{fig:t1_indloss} (b) shows the same $T_1$ data and fit with the qubit frequency on the x-axis.
The data in the range $f  \approx[4, 16.8]$ GHz (or $\Phi_\mathrm{ext}/\Phi_\mathrm{0}  \approx[-0.4, -0.01]$) does not agree well with either fit. 
There are clear frequency regions where the $T_1$ degrades before it improves again.
We conjecture that the presence of multiple TLS could be causing the large variability in $T_1$. 
Examples of TLS in the particular frequency (or flux) range is shown in the inset of  Fig.~\ref{fig:fullspec}(b). 
We do not see $T_1$ variability for $f<$ \qty{4}{GHz}. Further investigation is needed to determine the TLS spectral density of the TiN material.
Note that the Purcell loss limit is significantly higher than our measured relaxation times at $-0.5\Phi_\mathrm{0}$ or at $0\Phi_\mathrm{0}$ when setting the bath temperature to $T=$ \qty{174}{mK}.

\section{Qubit coherence vs flux analysis}
In Fig. 3(e) of the main text, we fit the $T_{2R}$ vs flux using the quadrature sum of the dephasing rate from the flux noise, Aharonov-Casher(AC) noise,  and noise associated with thermal excitations from the $\ket{e}$ to $\ket{f}$ state. We assume that these noise contributions are independent and Gaussian distributed \cite{randeria_dephasing_2024}.
\begin{equation} \label{eqtotalrate}
\frac{1}{T_{\varphi R}}=\Gamma_{\varphi R} =\sqrt{(\Gamma_{\varphi 
R}^{\Phi})^2+ (\Gamma_{\varphi R}^{\text{AC}})^2+(\Gamma_{\varphi R}^{\text{ef}})^2}
\end{equation}
To obtain the $T_{2R}$ vs flux fit (black line in Fig. 3(e) of the main text), we use
\begin{equation} \label{eqT2}
\frac{1}{T_{2 R}}=\frac{1}{2T_{1}} + \frac{1}{T_{\varphi R}}
\end{equation}
where $T_{\varphi R}$ is obtained from Eq.\ref{eqtotalrate}. We set $T_1$ to the values obtained in the $T_1$ vs flux fit near zero flux. However, $T_{2 R}\approx  T_{\varphi R}$ since $T_1>>T_{\varphi R}$. Below, we go through the analysis for the dephasing rate from both the flux noise and the Aharonov-Casher noise.
\subsection{Dephasing from flux noise}
The flux noise spectral density is generally given by
\begin{equation} \label{noise_spectrum_density}
S_{\Phi}(\omega) = A_{\Phi}^2\left(\frac{2\pi\times1\text{Hz}}{|\omega|}\right)
\end{equation}
where $A_{\Phi}$ is the flux noise amplitude. 
Following the theory in \cite{ithier_decoherence_2005}, the flux noise dephasing rate away from the flux sweet spots can be expressed as
\begin{equation} \label{eq2}
\Gamma_{\varphi R}^{\Phi} = \left|\frac{\partial\omega_{ge}}{\partial\Phi_{\mathrm{ext}}}\right|A_{\Phi}\sqrt{\ln\left(\frac{1}{\omega_{\mathrm{ir}}t}\right)+1}
\end{equation}
where $\omega_{\mathrm{ir}}$ is the low cut off frequency set by the duration of the Ramsey experiment and $t$ is the free evolution decay time.
The factor $\sqrt{\ln(1/\omega_{\mathrm{ir}}t+1)}$ is measurement specific, however, we treat it as a constant in our case, as it remains approximately constant over the range of $t$. 
The green line in Fig. 3(e) of the main text refers to the expected Ramsey coherence times, $T_{2R}^{\Phi}$,  from the flux noise contribution.  
The fitted flux noise amplitude, $A_{\Phi}$, is $1.41 \times 10^{-4} \Phi_0$/$\sqrt{\text{Hz}}$.

\subsection{Dephasing from Aharonov-Casher noise}
\begin{figure*}[ht]
\centering
    \includegraphics{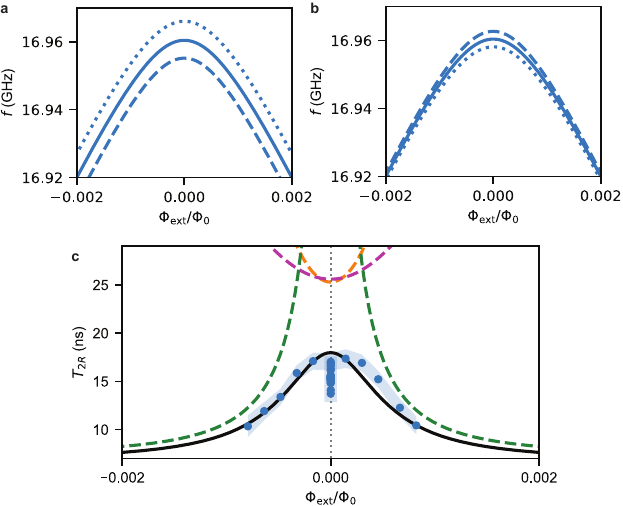}
    \caption{\label{fig:qubit_freq_Es} \textbf{Simulated qubit frequencies and  qubit coherence due to variations in $E_{s,\,2}$.} 
    (a) Qubit frequency near zero flux when $E_{s,\,1}$ is \qty{0.93}{GHz} (dashed line), \qty{1.03}{GHz} (solid line), and \qty{1.13}{GHz} (dotted line) while keeping $E_{s,\,2}$ constant at \qty{0.050}{GHz}.
    (b) Qubit frequency near zero flux when $E_{s,\,2}$ is \qty{0.045}{GHz} (dashed line), \qty{0.050}{GHz} (solid line), and \qty{0.055}{GHz} (dotted line) while keeping $E_{s,\,1}$ constant at \qty{1.03}{GHz}.
    (c) Ramsey coherence time $T_{2R}$ versus flux.
     The shaded region indicates the error from fitting the $T_{2R}$ data.
     The green line shows the expected flux dependence of $T_{2R}$ due to first-order flux noise dephasing with noise amplitude, $A_{\Phi} = 1.51\times 10^{-4} \Phi_0$/$\sqrt{\text{Hz}}$.
     The pink line shows the Ramsey coherence time resulting from Aharonov-Casher dephasing for variations in $E_{s,\,2}$ with noise amplitude, $A_{\text{AC2}} = 0.063E_{s,\,2}$/$\sqrt{\text{Hz}}$.
     The orange line represents the estimated coherence time from the dephasing due to proximity of the $\ket{f}$ state for a qubit temperature of \qty{17}{mK}.
     The black line is expected limit from the quadrature sum of all three models.
     The gray line is a guide for zero flux and the range of variations of $T_{2R}$ in time is shown by the data points at zero flux.
     Note that at zero flux, only the point with the highest $T_{2R}$ was included in the fit.}
\end{figure*}

Aharonov-Casher noise arises from fluctuations in the qubit frequency caused by variations in $E_{s,\,1}$ and $E_{s,\,2}$ over time. 
As described in the main text, the variation in the phase slip rates stems from changes in the distribution of offset charges along the wire. 
In this phenomenological model, we assume that the charge noise, which is typically modeled as a $1/f$ Gaussian distributed noise, also leads to a $1/f$ Gaussian distributed type noise in $E_{s,\,1}$ and  $E_{s,\,2}$. 
We can then express the dephasing rate from the AC noise in a similar manner to the dephasing rate from the flux noise. This is given by

\begin{equation} \label{AC}
\Gamma_{\varphi R}^{\text{AC}} = \left|\frac{\partial\omega_{ge}}{\partial E_{s,\,i}}\right|A_{\text{AC}}\sqrt{\ln\left(\frac{1}{\omega_{\mathrm{ir}}t}\right)+1}
\end{equation}
where $\partial\omega_{ge}/\partial E_{s,\,i}$ is the change in the qubit radian frequency with respect to a change in $ E_{s,\,1}$ or  $E_{s,\,2}$. 
Fig.~\ref{fig:qubit_freq_Es}(a) and (b) below show the qubit frequency near zero flux for different values of $ E_{s,\,1}$ and $E_{s,\,2}$, respectively. In our experiment, we cannot distinguish between dephasing resulting from $E_{s,\,1}$ and $E_{s,\,2}$. In the main text, we only consider the AC dephasing due to variations in $E_{s,\,1}$ since it is the leading order term in the Hamiltonian. This results in an AC noise amplitude, $A_{\text{AC}}$, of $ 0.026 E_{s,\,1} $/$\sqrt{\text{Hz}}$. 
It is possible that the charge noise could also lead to changes in $E_{s,\, 2}$. 
In Fig.~\ref{fig:qubit_freq_Es}(c), we consider the Ramsey coherence time from AC noise due to changes in $E_{s, \,2}$ only. 
From the fit, the AC noise amplitude is $A_{\text{AC2}} = 0.063E_{s,\,2}$/$\sqrt{\text{Hz}}$, while the flux noise amplitude increases slightly to $A_{\Phi}= 1.51 \times 10^{-4} \Phi_0\sqrt{\text{Hz}}$.

\subsection{Dephasing from excitations to the $\ket{f}$ state}

The $\ket{f}$ state is located $\sim\qty{70}{MHz}$ away from the $\ket{e}$ state at zero flux.
Thermally exciting to the $\ket{f}$ state from the $\ket{e}$ state during a measurement can therefore impact the dephasing rate of the qubit as described in \cite{ardati_using_2024}.
The decay rate from this effect is given by $\Gamma_{2 R}^{\text{ef}}=\Gamma_{\text{ef}}^\uparrow/2$.
From our inductive loss model, we can calculate $\Gamma_{\text{ef}}^\downarrow$, replacing $\bra f\hat\Phi\ket e$ and $S(\omega_{ef})$ in Eq.\ref{eqinductiveloss}.
We can use detailed balance to determine $\Gamma_{\text{ef}}^\uparrow$:

\begin{equation} \label{detailbalance}
\Gamma_{\text{ef}}^\uparrow = \exp\left(\frac{-\hbar\omega_{ef}}{k_BT}\right)\Gamma_{\text{ef}}^\downarrow,
\end{equation}

\noindent where $\omega_\mathrm{ef}$ is the $|e\rangle\rightarrow|f\rangle$ transition frequency and $T$ is the qubit temperature.
We expect that $|f\rangle$-state dephasing significantly contributes to the measured echo dephasing in our qubit since the slow noise associated with the fluctuating charges that cause Aharonov-Casher dephasing should be mitigated by the echo.
In Fig. 3(e) of the main text, we set $T$ to \qty{17}{mK}, which results in a coherence time of $\sim$\qty{25}{ns} at zero flux, in line with the measured $T_{2 E}$.
We fit the flux noise and AC noise amplitudes after finding the parameters associated with the $\ket{e}$ to $\ket{f}$ thermal transition.
The low extracted qubit temperature is in disagreement with our qubit temperatures extracted from histogram measurements ($T = 174\text{ mK}$). 
Thermal dephasing at this temperature would result in a significantly lower value; $T_{2 R}^{\text{ef}}$ would be at most 2.5 ns for the qubit temperatures that we measure at zero flux. 
A possible explanation for this discrepancy is that we may be overestimating the inductive loss in the circuit.
Future experiments where we vary the $|e\rangle-|f\rangle$ transition frequency will better elucidate the contribution of this dephasing mechanism.

\section{Operation of qubit at elevated temperatures}
\subsection{Extracting qubit temperature}
During the mixing chamber temperature sweep, we perform qubit histogram measurements where we extract $P_r$, the ratio of the probability of the qubit being in state $\ket{e}$ to the probability of the qubit being in state $\ket{g}$. 
We then use the Boltzmann distribution to calculate the qubit temperature.

\begin{equation} \label{boltzmann}
T_\mathrm{qubit} =\frac{-\hbar\omega_{ge}}{k_B \ln(P_r)}
\end{equation}

\noindent where $\omega_{ge}$ is the qubit radian frequency at zero flux. 
The qubit temperature from this calculation is shown as the orange points in Fig. 4(c) of the main text.

Furthermore, we can consider the possibility of thermally populating the $\ket{f}$ state.
In this calculation, we assume that the $\ket{f}$ state population may be included in the $\ket{e}$ state resonator response since we are not able to distinguish the $\ket{e}$ and $\ket{f}$ responses in the qubit histogram.
To obtain the qubit temperature with this assumption, we numerically solve the following equation for $T_\mathrm{qubit}$:
\begin{equation} \label{boltzmann}
\frac{P_e+P_f}{P_g}= P_r =\exp\left(\frac{-\hbar\omega_{ge}}{k_BT_\mathrm{qubit}}\right)+\exp\left(\frac{-\hbar\omega_{gf}}{k_BT_\mathrm{qubit}}\right)
\end{equation}
We account for this qubit temperature as the lower bound of the error bars in Fig.4(c) of the main text.
Note that for the first three temperature points, the uncertainty from the qubit histogram fit is larger, thus we use that error instead.
For the upper bound of the error bars for all the points, the uncertainty from the qubit histogram fit is used.

\subsection{Readout histogram dependence on resonator power}

\begin{figure*}[ht]
\centering
    \includegraphics{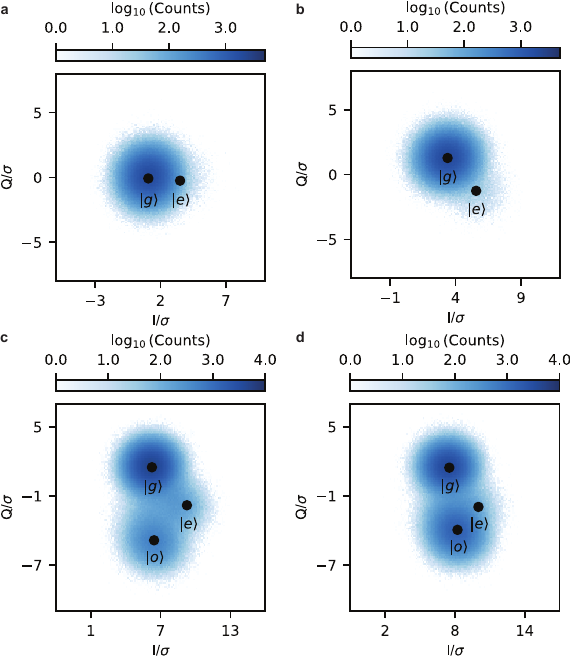}
    \caption{\label{fig:mist} \textbf{Measurement induced state transitions}
    (a)-(d) Qubit histogram measurements at base temperature with expected location of resonator response for qubit states using (a)  \qty{-128}{dBm}, (b) \qty{-124}{dBm}, (c) \qty{-119}{dBm}, (d) \qty{-116}{dBm}.
    The black dots represent the centers of the Gaussian distribution for each state's population, with the corresponding states labeled below it.}
\end{figure*}

To investigate the possibility of measurement-induced state transitions \cite{sank_measurement-induced_2016} affecting our temperature measurements,
we performed readout power dependence measurements of the qubit histograms (Fig.\ref{fig:mist}). 
We perform state assignments by first driving the qubit with a $X_{\pi/2}$ pulse to ensure population of both the ground and excited states (as shown in main text Fig.~2(b)).
For our qubit temperature measurements and the measurements shown in Fig.~\ref{fig:mist}, no pulse is performed prior to the measurement.
We varied the power of the readout signal sent to the device and observed if any additional states other than $\ket{g}$ and $\ket{e}$ appeared in the readout histograms.
The readout power sent to the resonator from Fig.\ref{fig:mist}(a)-(d) is \qty{-128}{dBm}, \qty{-124}{dBm},\qty{-119}{dBm}, \qty{-116}{dBm}, respectively.
A power of \qty{-128}{dBm} corresponds to roughly one photon in the resonator.
At \qty{-119}{dBm}, another state $\ket{o}$ starts to appear, while below this value, only $\ket{g}$ and $\ket{e}$ states are present.
Therefore, we readout below this power at \qty{-124}{dBm} (circles) and \qty{-120}{dBm} (crosses) in our qubit temperature measurements in Fig.4(c) of the main text.
Fitting the histograms in Fig.\ref{fig:mist}(a) and (b) yields qubit temperatures of \qty{153}{mK} and \qty{174}{mK} respectively, which is within the uncertainty range of the fit.

\subsection{Temperature dependence of qubit lifetime }
To model the decrease in $T_1$ with increasing mixing chamber temperature, we assume that $Q_\mathrm{ind}$ changes to reflect the increase in the number of quasiparticles \cite{amin_loss_2022}: 

\begin{equation} \label{Qi_temp}
\frac{1}{Q_{\mathrm{ind}}(T) }=2\alpha\sqrt{\frac{k_BT}{\pi \hbar \omega}} \exp\left({\frac{-\Delta_0}{k_B T}}\right)+\frac{1}{Q_{\mathrm{ind}}^{\text{initial}}}
\end{equation}

\noindent We assume that the bath temperature remains the calculated qubit temperature at each mixing chamber temperature point in Eq.\ref{Qi_temp} and Eq.\ref{eq_noisespectal}. 
We let $Q_\mathrm{ind}^{\text{initial}} = 5.0\times10^4$, which is a slight adjustment from the fitted value near the zero flux flux sweep, to match the initial $T_1 = $ \qty{66}{\mu s} data point during the temperature sweep.
\section{Device B measurements}

\subsection{Qubit spectroscopy and time domain measurements of Device B at base mixing chamber temperature}
\begin{figure*}[ht]
\centering
    \includegraphics{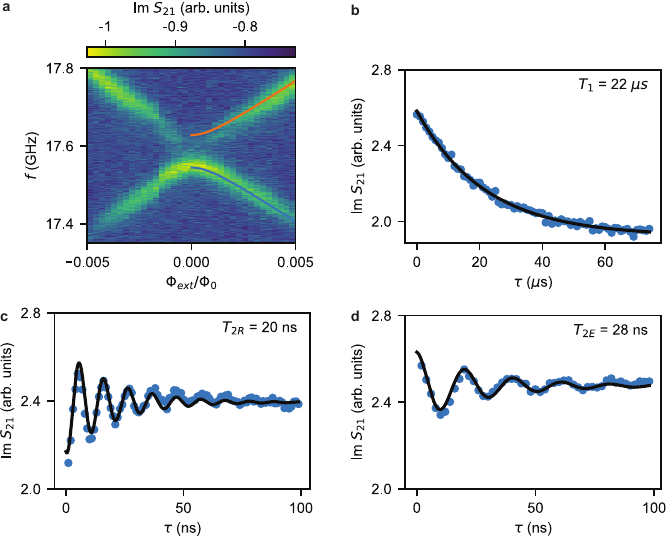}
    \caption{\label{fig:dBfit} \textbf{Device B qubit spectroscopy and time domain measurements}
    (a) Two-tone spectroscopy near zero flux.
    The fit to the $\ket{g}-\ket{e}$ transition (blue line) and $\ket{g}-\ket{f}$  transition (orange line) is shown on the right half of the data. 
    (b-d) Time domain measurements taken at zero flux measuring (b) the relaxation time, $T_1$, (b) the Ramsey coherence time, $T_{2R}$, and (c) the coherence time with the application of an echo pulse, $T_{2E}$.
    }
\end{figure*}

To evaluate the reproducibility of the results, we performed measurements on another device (B).
Although Device B is located in a different chip, both devices were designed to have the same parameters and have gone through the same fabrication process.
In Fig.~\ref{fig:dBfit}(a), we plot the two-tone spectroscopy data near zero flux for Device B.
We were unable to perform two-tone spectroscopy close to the half-flux point due to a reduction in visibility of the resonator signal.
Thus, we only fit the qubit Hamiltonian to the data near zero flux, which leads to a larger uncertainty in the phase slip rates compared to Device A.
From the fit, we extract $E_L = 35.1\pm0.1$ GHz, $E_{s,\, 1} =0.9\pm0.9$ GHz, and $E_{s,\,2} =0.04\pm0.04 $ GHz.
The difference in $E_L$, and correspondingly the qubit frequency, between Device A and B is only $\sim 3\%$, which demonstrates the predictability of our qubits.

In Fig.~\ref{fig:dBfit} (b), we present the results of Device B's time domain measurements at zero flux. We find that the qubit lifetime with $T_1 = $ \qty{22}{\mu s}  is lower than Device A's lifetime at zero flux, however, the coherence time is slightly higher with $T_{2R} = $ \qty{20}{ns}.

\subsection{Device B time domain measurements at elevated temperature}
\begin{figure*}[ht]
\centering
    \includegraphics{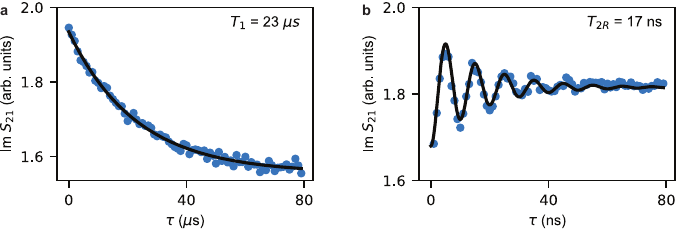}
    \caption{\label{fig:dB200mK} \textbf{Device B time domain measurements at \qty{200}{mK}}
    (a) Measurement of relaxation time, $T_1$.
    (b) Measurement of Ramsey coherence time, $T_{2R}$.
    Neither $T_1$ nor $T_{2R}$ changes significantly compared to measurements conducted at base temperature.
    }
\end{figure*}

We also investigate the operation of Device B at elevated temperatures. 
We raise the temperature of the mixing chamber to \qty{200}{mK} and perform time-domain measurements. 
We find that the qubit lifetime and coherence times are relatively unaffected at this temperature (See Fig.~\ref{fig:dB200mK}).

\newpage

\bibliography{supplement1}   
\nocite{*}